# Enhanced structural correlations accelerate diffusion in charge-stabilized colloidal suspensions

Ekkehard Overbeck and Christian Sinn[*]

Institut für Physik der Johannes-Gutenberg-Universität, 55099 Mainz, Germany

Martin Watzlawek

Institut für Theoretische Physik II der Heinrich-Heine-Universität, 40225 Düsseldorf, Germany

---

[*] author for correspondence
http://www.uni-mainz.de/FB/Physik/Kolloide/

Theoretical calculations for colloidal charge-stabilized suspensions and hard sphere suspensions show that hydrodynamic interactions yield a qualitatively different particle concentration dependence of the short-time self-diffusion coefficient. The effect, however, is numerically small and hardly accessible by conventional light scattering experiments. Applying multiple-scattering decorrelation equipment and a careful data analysis we show that the theoretical prediction for charged particles is in agreement with our experimental results from aqueous polystyrene latex suspensions.

Colloidal systems have attracted considerable attention recently, as many aspects of structure and dynamics of condensed matter can be modeled theoretically and are accessible experimentally. Investigations are dedicated, *e.g.*, to the caging effect of direct neighbor interaction and its implication for properties like the suspension's structure factor or the diffusion tensor [1,2,3]. In experimental studies, mainly two kind of colloidal suspensions are used. Uncharged sterically stabilized particles with a thin hairy polymer layer on the surface resemble very closely a hard sphere interaction potential. On the other hand, the interaction potential of particles with some surface charges on a compact core can be well described by DLVO theory, that is by a Debye-Hückel-type long-range electrostatic repulsion with an underlying short-range van-der-Waals attraction. With increasing understanding of physical effects due to the (direct) interaction potential also indirect, namely hydrodynamic interactions (HI), became an attractive research topic. These are difficult to describe theoretically because of the necessity to tackle many-body interactions [1,2,3].

In this contribution we focus on the influence of HI on the (translational) short-time self-diffusion coefficient $D$, which is a key feature of colloidal dynamics, because collective diffusion or long-time memory effects can only be determined if the short-time self-diffusion of the particles is known in detail [1,2]. For small particle volume fractions $\phi$, $D$ can be written as a (hydrodynamic) virial expansion

$$D/D_0 = 1 + H_1 \cdot \phi + H_2 \cdot \phi^2 + O(\phi^3), \tag{1}$$

where $H_1$ incorporates two-body HI and is given by

$$H_1 = \frac{1}{R^3} \int_0^\infty r^2 g(r) M_{11}(r) \, \mathrm{d}r. \tag{2}$$

$g(r)$ is the particle pair correlation function and $M_{11}(r)$ is a hydrodynamic mobility function, both of which depend on the distance $r$ of the particles with radius $R$ only. $H_2$ accounts for three-body HI [2,4], $D_0$ is the single-particle diffusion coefficient.

For hard spheres, the well-established result is $H_1 = -1.831$ and $H_2 = +0.71$ [5,6,4]. Hard sphere suspensions can be considered to be dilute both with respect to particle hydrodynamics and microstructure as well. Therefore, low-density approximations for the hard sphere particle correlation functions can be applied in the calculation of $H_1$ and $H_2$ at small $\phi$. In contrast, charge-stabilized suspensions can be regarded as dilute with respect to the hydrodynamics only, as $g(r)$ exhibits a pronounced structure incorporating a correlation hole even at small $\phi$ [4,7]. Its extension depends strongly on $\phi$, but also on the concentration of added electrolyte, $c_s$. Due to the strong particle repulsion, the main peak of $g(r)$ is located at the mean particle distance $d = R[3\phi/(4\pi)]^{-1/3}$. For a rough estimate of how this affects the self-diffusion coefficient, we approximate the pair correlation function by a unit step-function $g(r) = \Theta(r-d)$, thus neglecting structural details, but retaining the important correlation hole in a simplified fashion. If we keep only the leading term of the tabulated far-field expansions for the mobility function, $M_{11}(r) = -(15R^4)/(4r^4) + O(r^{-6})$ [8], we obtain from Eqs.(1) and (2) by taking only two-body HI into account,

$$D/D_0 = 1 - a \cdot \phi^b, \qquad (3)$$

with $a = 2.33$ and $b = 4/3$. Within these approximations, we thus expect a transition from a linear to an algebraic concentration dependence of $D/D_0$ with increasing structural correlations. This qualitative different behavior for charge-stabilized systems as compared to hard spheres still persists if two-body HI is completely taken into account (up to $O[r^{-20}]$) and three-body HI to the leading order [4]. In Fig. 1, we show the corresponding numerical results for $D/D_0$ as a function of $\phi$ for different $c_s$. The numerical procedure for the calculation has been described in Ref. [4]. The exact results for $a$ and $b$, respectively, are rather insensitive to the detailed form of the potential, and to the closure relation used to calculate $g(r)$ via integral equation theory, as long as the repulsion of the particles is strong enough to keep the relation $d \propto \phi^{-1/3}$ valid [4].

Up to a volume fraction of approximately $\phi = 0.06$, the curves can be fitted to a power law according to Eq.(3). For the salt-free case, *e.g.*, we obtain $a = 2.37$ and $b = 1.28$, very close to the leading behavior derived above. With increasing $c_s$, that is increasing screening of the repulsion, $b$ rapidly approaches the hard sphere result $b = 1$ due to decreasing structural correlations between the particles [4,7]. We emphasize that, as visible in Fig. 1, structural correlations in charge-stabilized suspensions increase the diffusion coefficient as compared to hard spheres owing to decreasing hindering by HI with increasing particle repulsion.

The experimental observation of this effect, however, seems to be extremely difficult. HI become important above, say, $\phi = 0.01$ only (*cf.* Fig. 1). At these particle concentrations, multiple scattering contributions in general strongly prohibit optical investigations. One possibility, therefore, is to investigate near-isorefractive systems, thereby strongly reducing the particle scattering cross section. This approach has been successfully exploited in sedimentation experiments, where the general features of the theory could be confirmed, but certain deviations prevailed possibly due to incomplete knowledge of the pair potential [9].

In this contribution we employ an aqueous polystyrene latex suspension, where the pair potential can be very effectively controlled by fixing the amount of added salt. To be specific, $\phi$ and $c_s$ can be precisely determined by measuring the suspension's low-frequency conductivity [10]. The conductivity can be reproducibly adjusted in a closed tubing system between the minimum value, given by the self-dissociation of water plus the dissociation of the particle charges, and the maximum value set by aggregation of the particles due to the complete screening of the electrostatic repulsion.

Using this well characterized system, however, we face two experimental challenges. First, multiple scattering will be strong due to the difference in refractive index between particles and water of approximately $\Delta n = 0.3$. In addition, routine measurements of diffusion coefficients by dynamic light scattering (DLS) are performed with an accuracy of about ±5% only [11]. This error covers the whole expected effect of the difference between hard spheres and charged spheres even at zero foreign salt content, as visible in Fig. 1.

Nevertheless, we shall determine in this contribution the short-time self-diffusion coefficient in concentrated polystyrene latex suspensions with high precision. We decorrelate multiple scattering contributions by taking advantage of a novel DLS experiment introduced recently [12], and perform a careful statistical data analysis. We shall show that the data we obtain exhibit qualitatively as well as quantitatively the effect discussed above.

DLS cross-correlation functions can be shown to be essentially free from multiple scattering contributions when they are determined from intensity fluctuations measured by two geometrically different scattering experiments with common scattering volume and equal scattering vector $\vec{q}$ [13]. One feature of our experiment, in particular, is the long wavelength used ($\lambda = 788.7$ nm), which allows the investigation of strongly turbid samples, as the scattering cross section is reduced. In addition, by using thin rectangular cuvettes with short optical light paths ≤ 1 mm, we are able to still further reduce the amount of multiply scattered light. Thus, even undiluted cow milk could be investigated by DLS [14].

Self-diffusion is accessible in DLS experiments on interacting colloidal samples if the scattering vector $q = (4\pi n/\lambda)\sin(\theta/2)$ is significantly larger than the position of the maximum of the structure factor, $q_{max} \approx 2\pi/d$; $n$ is the refractive index of the suspending medium and $\theta$ the scattering angle. In the short-time limit, one then measures single-particle dynamics, which should not be influenced by direct next-neighbor interactions, *i.e.* the caging effect. We use polystyrene spheres with a nominal radius of $R = 135$ nm, determined by electron microscopy. Owing to their strong repulsion at $c_s = 50$ µmol/dm$^3$, the first maximum of the fluid structure factor is located at a scattering angle of approximately 45° with $\phi = 0.02$. We determine the correlation functions at ten different scattering angles between 120° and 140°, where we expect any oscillations of the structure factor due to structural order to be decayed, in accordance with experimental observation. We measure the cross-correlation functions of the individual suspensions under investigation by averaging 30 single correlator runs lasting 100 s each. A cumulant expansion of third order is fitted to the correlation function between the 10th and 90th ALV-5000/E correlator channel. From the ten resultant first cumulants we calculate the mean diffusion coefficient and its standard deviation, which could be reduced this way to typically 1%, which is significantly smaller than in usual DLS measurements.

The suspensions under study were prepared as follows. We add mixed bed ion exchange resin to the raw colloidal suspension. The mixture was left to stand several days with occasional stirring. This stock suspension was filtered using 5 µm pore size filters and subsequently filled into a closed tubing system with a peristaltic pump, where the conductivity of the suspension can be controlled by the resin, through which the suspension is pumped or bypassed [10]. We use the minimal conductivity as a measure for the particle concentration in the circuit. Calibration was performed by measuring the minimal conductivity of nine suspensions of different $\phi$ within the circuit. The polystyrene content was determined independently by weighing a carefully dried amount of the suspensions. Volume fractions can now be calculated from the density of the polystyrene particles in the suspension. We adopt the value given by the manufacturer ($\rho$ = 1.055 g/cm$^3$), which results to be the main source of error in $\phi$. From these calibration measurements we obtain an effective particle charge of $Z_{\text{eff}}$ = 2500 if we assume a particle mobility of $\mu_P = 10^{-7}$ m$^2$/(Vs). The suspensions tend to crystallize at the minimal conductivity that corresponds to no added salt. To maintain the fluid state, we tried to fix the conductivity at a non-vanishing foreign salt concentration, say $c_s$ = 50 µmol/dm$^3$. Experimentally, however, this is not easy to guarantee, as $\phi$ and $c_s$ are altered concomitantly. We therefore determine the correlation function at constant conductivity and remove the added salt with the ion exchange resin afterwards. From the measured minimal conductivity we calculate $\phi$, from the conductivity difference we calculate $c_s$. We then add salty water to the circuit in order to change $\phi$ and to increase $c_s$ again. Seven suspensions with different $\phi$ were investigated employing this scheme; $c_s$ could be kept within $c_s$ = [30..100] µmol/dm$^3$ during these measurements.

In order to fit the data to Eq.(1) or Eq.(3), the corresponding reference value for infinite dilution, $D_0$, is needed. Usually, one would expect this to be the Stokes-Einstein diffusion coefficient $D_{\text{SE}} = k_B T/(6\pi\eta R)$, with $\eta$ the viscosity of the solvent, determined in a strongly diluted suspension with high salt content ($c_s$ = 500 µmol/dm$^3$); $D_{\text{SE}}$ = (1.536±0.009) µm$^2$/s. However, for our data, $D_0$ has to be determined at the same Debye screening length $\kappa^{-1}$ as the measurements at high particle concentrations; $\kappa^2 = L_B\left[3Z^*\phi/(4\pi R^3) + 2N_L c_s\right]$.

$L_B = e^2/(\varepsilon\varepsilon_0 k_B T)$ is the Bjerum length at the solvent dielectric number $\varepsilon$, $Z^*$ is an effective charge, in general different to $Z_{\text{eff}}$ obtained before [3]. We observed experimentally that $D_0$ differs from $D_{\text{SE}}$, probably owing to electroviscous effects. We determined the diffusion coefficient of three semi-dilute samples ($\phi$ = 0.009, 0.006, and 0.002) as a function of $\kappa$ in that conductivity range, where also the higher concentrated samples were investigated. We call these samples semi-dilute as HI has negligible influence on the data. We found that all data points fell onto a common curve. Starting with $D_{SE}$ at $\kappa R = 6$, $D_0$ decreases monotonically by 10% up to $\kappa R = 2$. Following this observation, we determine the wanted $D_0$ for a suspension at high $\phi$ by measuring $\kappa$ and taking the corresponding $D_0$ value from the common curve. We shall discuss these observations in detail in a future paper. From the mere observation of a master curve, we presume to measure single-particle electroviscous effects at low $\phi$. At the highest volume fraction investigated ($\phi$ = 0.06), we have a Debye screening length of $\kappa^{-1}$ = 40 nm and a mean particle distance $d \approx 550$ nm. As $\kappa^{-1} \ll d$, we conclude that our extrapolation is valid even for those higher volume fractions. However, we have no proof for this assertion.

The result of the procedure described before is shown in Fig. 2. Our data points are clearly located above the hard sphere result. This at first *qualitatively* confirms the expectation that diffusion is less hindered by HI for charged spheres as compared to hard spheres, leading to a faster diffusion due to enhanced structural correlations. The error bars represent the standard deviation of the measured diffusion coefficients and an arbitrarily chosen 5% uncertainty in particle density, respectively.

Second, our data can be fitted to the expected power-law behavior of Eq.(3). The resulting fitting parameters $a = 1.7\pm1.1$ and $b = 1.0\pm0.2$ agree *quantitatively* with a fit of the data from the numerical calculation for $c_s = 50$ µmol/dm$^3$ and $\phi \leq 0.06$ ($a = 1.54$, $b = 1.09$). This is also visible in Fig. 2, where the numerical result has been included.

To summarize, we have shown experimentally for the very first time that charge-stabilized colloidal particles exhibit a different concentration dependence of the short-time self-diffusion coefficient than hard sphere suspensions. This result could only be obtained by employing a novel cross-correlation DLS technique for the decorrelation of multiple scattering and by taking into account the electroviscous effect, which is found to be not negligible in the range $\kappa R = [2..6]$. We note that our result is not in contradiction to the observations of Zahn et al. [15], who failed to observe a concentration dependence of the short-time self-diffusion coefficient upon investigating long-time caging effects, probably owing to their smaller experimental resolution.

We finish this contribution with some remarks concerning the DLVO interaction potential. Our numerical calculations are obtained assuming a fixed "effective" particle charge number $Z^*$ for the calculation of the static structure factor $S(q)$, which could not be determined experimentally with sufficient accuracy because of the strong multiple scattering of the suspensions. Therefore, as an estimate for our particles, we use $Z^* = 1000$, which may be justified from the observation that our measured phase boundary ($\phi = 0.01$ at $c_s = 0$ μmol/dm$^3$) coincides with the Hansen-Verlet criterion applied to the theoretical structure factors. The numerical results are known not to change noticeably for larger $Z^*$ [4]. In contrast, we use $Z_{eff} = 2500 > Z^*$ for the calculation of our $\kappa$ data from conductivity measurements. We note that the effective charges of the DLVO interaction potential are generally found to depend on the determination method (see, *e.g.*, Ref. [10]) and do not necessarily agree.

The authors are grateful for numerous stimulating discussions with Gerhard Nägele and his introduction into the subject. We are indebted to thank Thomas Palberg for his interest in this work and many helpful suggestions. This work was supported by DFG via SFB 262 and 513.

**Figures**

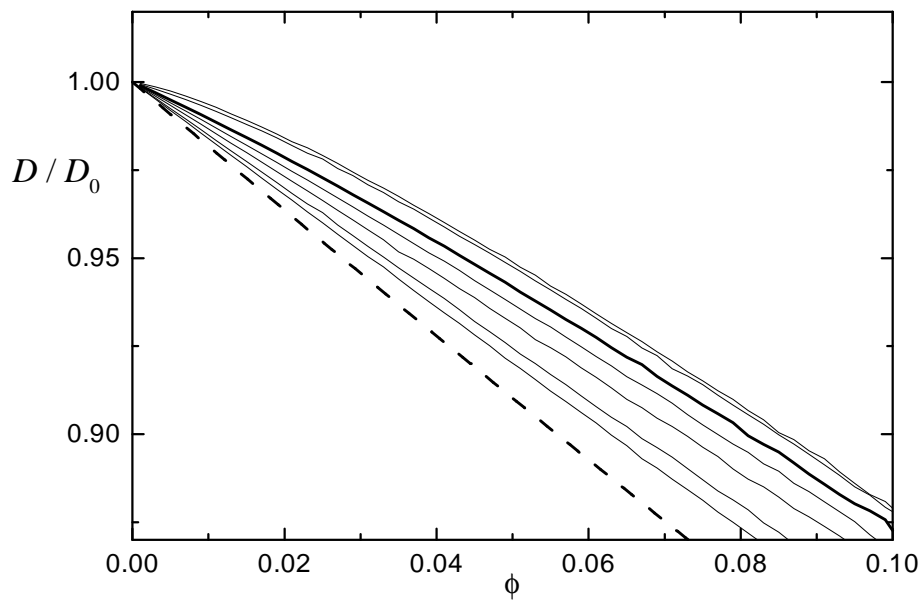

**Fig. 1:**

Numerical results for the short-time self-diffusion coefficient as a function of the particle volume fraction. Parameter is the foreign salt concentration $c_s$, from top to bottom: $c_s$ / µmol/dm$^3$ = {0, 10, **50**, 100, 200, 500 1000}, broken line: theoretical result for hard spheres. The system parameters are $Z^* = 1000$, $R = 135$ nm, $\varepsilon = 81.0$, and $T = 293.2$ K, resembling very closely the experimental situation.

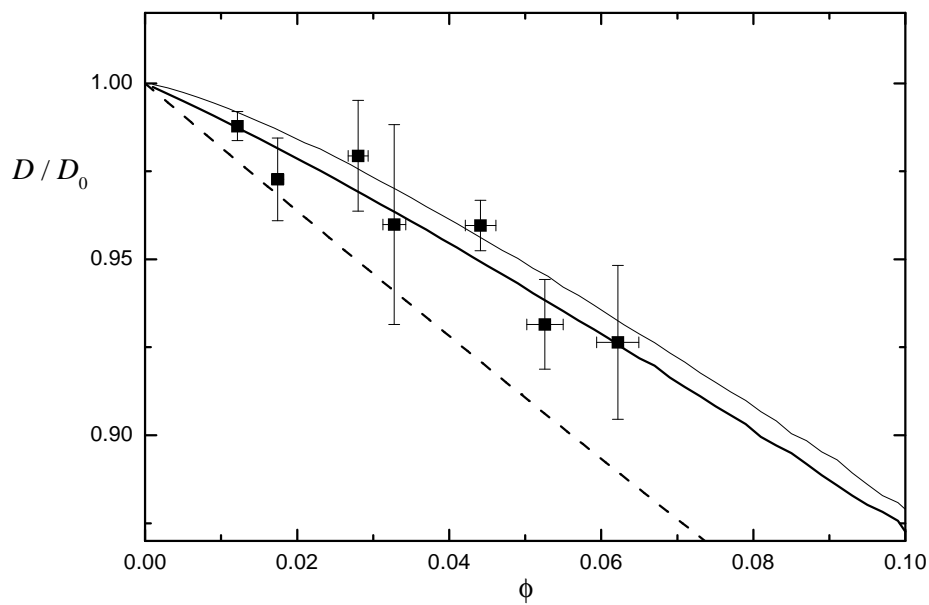

**Fig. 2:**

DLS data for the normalized short-time self-diffusion coefficient as a function of the particle volume fraction. Broken line: theoretical result for hard spheres, thick solid line: theoretical result for $c_s = 50$ μmol/dm$^3$, thin solid line: theoretical result for $c_s = 0$ μmol/dm$^3$.

**References**

1    P. N. Pusey, Colloidal suspensions, in *Liquids, Freezing and Glass Transition*, proceedings of the Les Houch sommer school 1989, edited by J. P. Hansen, D. Levesque, and J. Zinn-Justin (North Holland, Amsterdam 1991), p. 763.

2    J. K. G. Dhont, *An Introduction to Dynamics of Colloids*, (Elsevier, Amsterdam 1996).

3    G. Nägele, Physics Reports **272,** 215 (1996).

4    M. Watzlawek and G. Nägele, Phys. Rev. E **56,** 1258 (1997).

5    G. K. Batchelor, J. Fluid Mech. **74,** 1 (1976).

6    C. W. J. Beenakker and P. Mazur, Physica A **120,** 388 (1983).

7    G. Nägele, M. Watzlawek, and R. Klein, Progr. Colloid & Polym. Sci. **104,** 31 (1997).

8    D. J. Jeffrey and Y. Onishi, J. Fluid Mech. **139,** 261 (1984).

9    D. M. E. Thies-Weesie, A. P. Philipse, G. Nägele, B. Mandl, and R. Klein, J. Colloid Interf. Sci. **176,** 43 (1995).

10   M. Evers, N. Garbow, D. Hessinger, and T. Palberg, Phys. Rev. E **57,** 6774 (1998).

11   Compare, *e.g.*, A. van Veluwen, H. N. W. Lekkerkerker, C. G. de Kruif, and A. Vrij, J. Chem. Phys. **87,** 4873 (1987), for a careful determination of diffusion coefficients, which is comparable in scope with the present study.

12   E. Overbeck and C. Sinn, J. Modern Optics **46,** 303 (1999).

13    K. Schätzel, J. Modern Optics **38,** 1849 (1991).

14    C. Sinn, R. Niehüser, E. Overbeck, and T. Palberg,
      Progr. Colloid & Polym. Sci. **110,** 8 (1998).

15    K. Zahn, J. M. Méndez-Alcaraz, and G. Maret, Phys. Rev. Lett. **79,** 175 (1997).